\documentstyle[epsf,epsfig,12pt,ssi]{article}

\vspace{0.2cm}

\begin{document}
\title{Final Results on Heavy Quarks \linebreak at LEP and SLD}
\vspace*{2cm}
\author{Achille Stocchi \\ 
Laboratoire de l'Acc\'el\'erateur Lin\'eaire, \\ 
        IN2P3-CNRS et Universit\'e de Paris-Sud, BP34, F-91898 Orsay Cedex, France.\\[0.4cm]
}

\vspace{2cm}
\maketitle

\begin{abstract}
In the last decade, the LEP and SLD experiments played a central role 
in the study of B hadrons (hadrons containing a $b$ quark).
 New B hadrons have been observed ($B^0_s$, $\Lambda_b$, $\Xi_b$ and $B^{**}$) 
and their production and decay properties have been measured. 
In this paper we will focus on measurements of the 
CKM matrix  elements~: $|V_{cb}|$, $|V_{ub}|$, $|V_{td}|$ and $|V_{ts}|$.
We will show how all these measurements, together with theoretical developments, 
have significantly improved our knowledge on the flavour sector of the Standard Model.
\end{abstract}

\section{Introduction}
\label{sec:intro}
B physics studies are exploiting a unique laboratory for testing 
the Standard Model in the fermion sector, for studying the QCD 
in the non-perturbative regime and for searching for New Physics 
through virtual processes.\\
In the last decade, the LEP and SLD experiments played an important role 
in the study of B hadrons. At the start of the LEP and SLC accelerator in 1989, 
only the $B_d$ and the $B^+$ hadrons were known and their properties were under study.
 New weakly decaying B hadrons have been observed ($B^0_s$, $\Lambda_b$, $\Xi_b$) for 
the first time and their production and decay properties have been measured.
New strongly decaying hadrons, the orbitally (L=1) excited B
($B^{**}$) mesons have been also observed and their mass and 
production rates measured. \\
In this paper we will focus on the measurements 
of the CKM matrix elements~: $V_{cb}$ and $V_{ub}$ through B decays 
and $V_{td}$ and $V_{ts}$ using $B^0-\bar{B}^0$ oscillations. 
On the other hand many additional measurements on B meson properties 
(mass, branching fractions, lifetimes...) are necessary to constrain 
the Heavy Quark theories (Operator Product Expansion (OPE) /Heavy Quark Effective Theory (HQET) 
/Lattice QCD (LQCD)) to allow for precise extraction of the CKM parameters.  
We finally show how these measurements constrain the Standard Model in the fermion sector, 
through the determination of the unitarity triangle parameters.\\
In this paper we try to compare the LEP/SLD results with those obtained from other
collaborations (CLEO at Cornell, CDF at TeVatron and the asymmetric B-factories: 
BaBar and Belle) and to present,
when available, the world average result. A detailed description of the results and of 
the averaging techniques can be found in \cite{ref:bphys,ref:stocchi}.

\section{B physics at the $Z^0$}
At the $Z^0$ resonance, B hadrons are produced from the coupling of the 
$Z^0$ to a $b \bar{b}$ quark pair. The production cross section is of $\sim$ 6 nb, 
which is five times larger than at the $\Upsilon(4S)$.
Because of the specific (V-A) behaviour of the electroweak coupling at the $Z^0$ pole,  
hadronic events account for about  70$\%$ of the total production rate; among 
these, the fraction of $b \overline{b}$ events is $\sim 22\%$.
 Because of the energy available only $B^+$ and $B_d^0$ mesons can be 
 produced at the $\Upsilon(4S)$. 
The  B particles are produced almost at rest (the average momentum is of about \linebreak 350 MeV/c), 
with no accompanying additional hadrons, and the decay products of the two 
B particles are spread isotropically over the space. 
 At the Z pole, the primary $b \bar{b}$ pair, picks up from the 
vacuum other quark-antiquarks pairs and hadronizes into B hadrons plus 
few other particles.
 Therefore not only  $B^{\pm}$ and $B_d$ mesons are produced, but also 
 $B_s^0$ mesons or $B$ baryons can be present in 
the final state. 
The $b$ and the $\bar{b}$ hadronize almost independently. $b$ quarks 
fragment differently from light quarks, because of their  high mass as
compared with $\Lambda_{QCD}$. B hadrons carry, on average,  
about 70$\%$ of the available beam energy, whereas the rest of the energy is 
distributed among the other fragmenting particles. As a consequence, 
the two B hadrons fly in opposite directions and their decay products 
form jets situated in two opposite hemispheres. \\
The hard fragmentation and the long lifetime of the b quark 
make that the flight distance of a B hadron at the Z pole, defined 
as $L = \gamma \beta c \tau$, on average of the order of 3 mm. 
As decay products have a mean charged multiplicity of 5
\footnote{On average there are as many particles originating from $b$-quark
fragmentation and from B decay.}, 
it was possible to tag B hadrons using a lifetime tag.  

\begin{figure}[htbp!]
\begin{center}
\includegraphics[angle=-90,width=15cm]{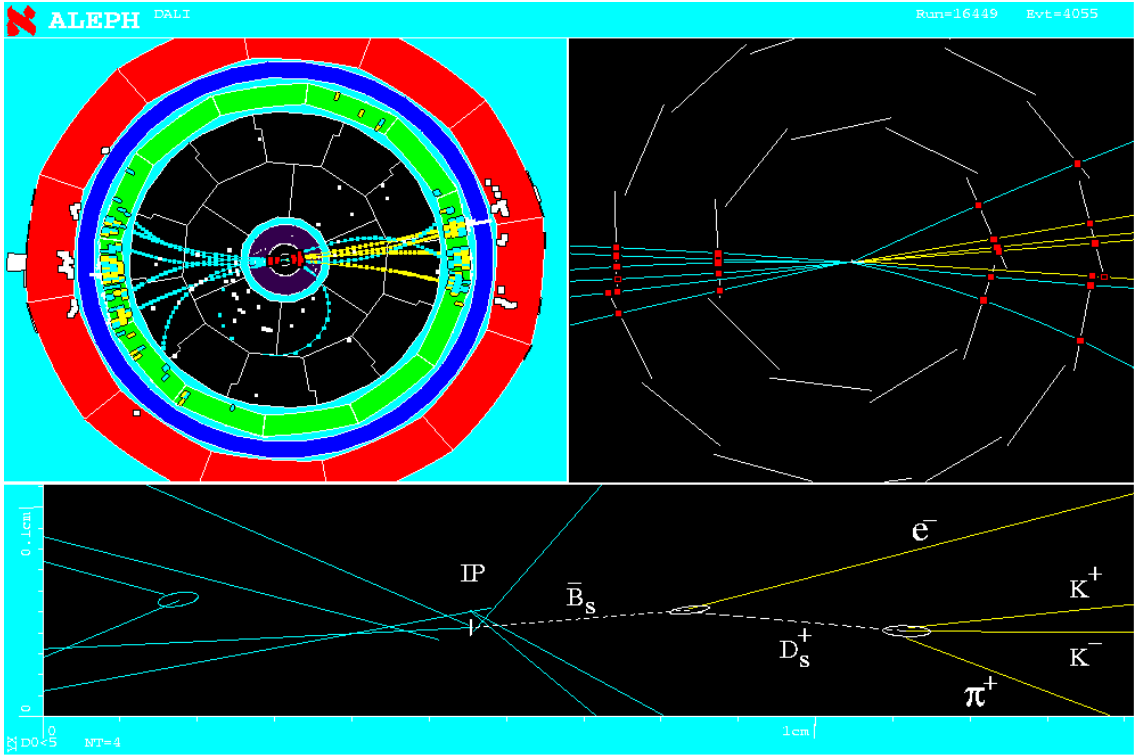}
\caption[]{\it{A LEP $b \bar{b}$ event. In the upper part, the ALEPH detector and a 
zoom on the charged tracks seen by the silicon detectors are displayed. In the lower part 
the reconstructed event is shown. The event is constituted of two jets
which define two separate hemispheres. In one of these hemispheres a $\bar{B}_s^0$
decays semileptonically~: $\bar{B}^0_s \rightarrow D_s^+ e^- \bar{\nu}_e X$ (secondary
vertex), followed by the decay : $ D_s^+ \rightarrow K^+ K^- \pi^+$ (tertiary vertex).
The primary vertex (marked with IP) is also shown.}}
\label{fig:aleph_event}
\end{center}
\end{figure}

Most of the precision measurements in B physics performed at LEP/SLC, Tevatron and B factories, 
would not have been possible without the development of Silicon micro vertex detectors. 
In practice the averaged flight distance of the B hadrons becomes measurable thanks 
to the precision of silicon detectors, located as close as possible to the beam interaction point.
 These detectors determine with a precision better than 10 $\mu m$, the position of a charged particle 
trajectory. In particular the separation between $b$ quarks and other quarks 
is mainly based on the use of vertex detectors. Charged particles produced 
at the B vertex (secondary vertex) can be separated from those produced 
at the interaction point (primary vertex) using the precise tracking information.
In spite of the relatively small statistics collected by the SLD experiment,
 it gave very important and competitive contributions to B physics, because of 
 its silicon vertex detectors, which is located very close 
to the interaction point.
A typical LEP $b\bar{b}$ event is shown in Figure \ref{fig:aleph_event}.
\
\
\begin{figure}[htbp!]
\begin{center}
\includegraphics[width=11cm]{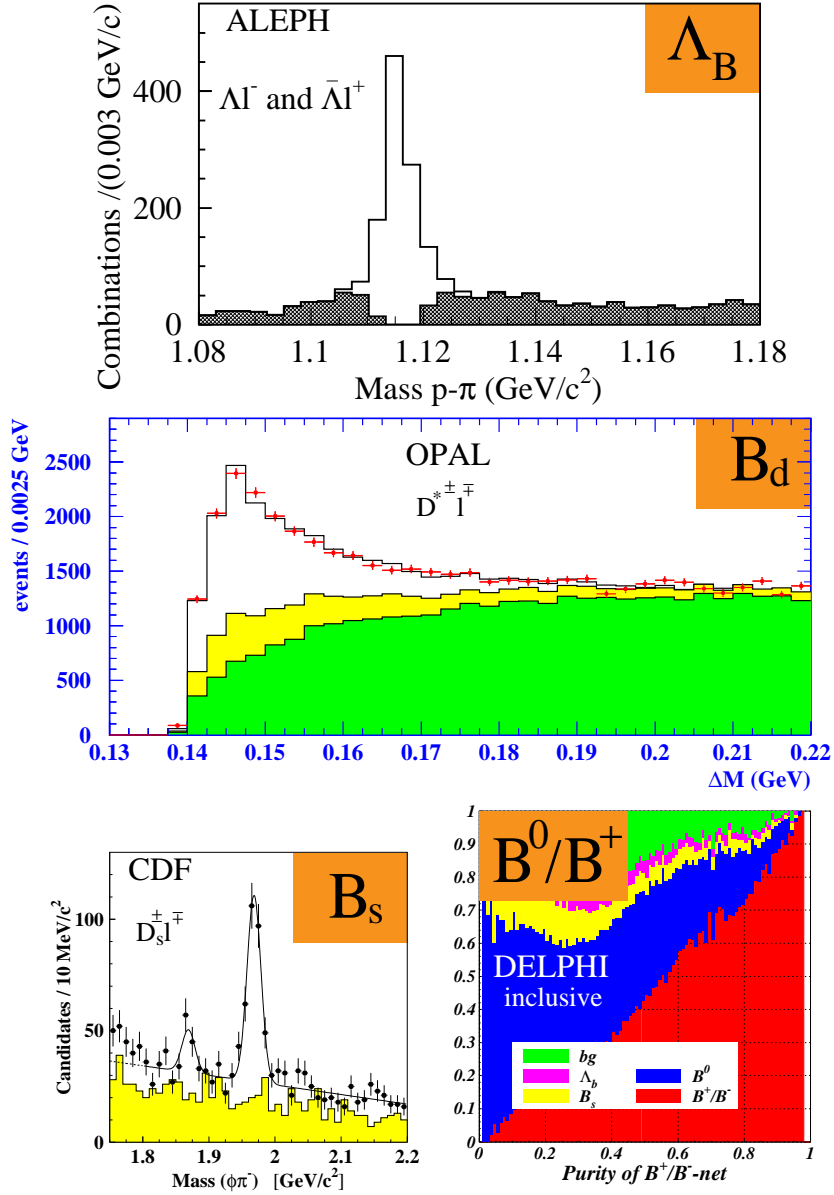}
\caption[]{\it{The three plots from top-left to bottom-left show the 
invariant mass spectra of ${\Lambda}$, ${((D^0 \pi)-D^0)}$ and ${D_s}$ which 
are obtained in correlation with 
an opposite sign lepton. These events are attributed mainly to the semileptonic decays 
of ${\Lambda_B}$, ${B_d^0}$ and ${B_s^0}$ hadrons, respectively. 
The bottom-right figure shows the possibility of distinguishing the charged and
neutral B mesons based on inclusive techniques.}}
\label{fig:4figures}
\end{center}
\end{figure}
Because of the large B mass, B hadrons are expected to decay 
into several decay modes with branching ratio of the order 
of a per mil.
\newpage
\noindent
 According to the registered statistics, at LEP, 
inclusive or semi-exclusive $b$-hadron
decays had to be studied in place of exclusive channels for which 
very few events are expected\footnote{with the final LEP statistics, 
B rare decays with branching fraction of the order of a few 
10$^{-5}$ could be accessed.}.\\
Semileptonic decays benefit of a large branching ratio ( of the order 
of 10$\%$ ) and of clean and easily distinguishable final states. 
Semileptonic decays allow also to distinguish between different types 
of $b$ hadrons, by reconstructing charmed hadrons.
As an example, a $\Lambda_c^+$ accompanied by a lepton with negative 
electric charge, in a jet, signs a $b$-baryon.
For baryons, it is not even necessary to completely reconstruct the
 $\Lambda_c^+$ charmed baryon, correlations as $p \ell^-$ or 
$\Lambda \ell^-$ are sufficient.
Similarly, $D_s^+ \ell^-$ or $D^* \ell^-$ events in a jet, provide 
event samples enriched in $\bar{B}^0_s$ and  $\bar{B}^0_d$ mesons 
respectively. \\
An overview  of the signals used to study these new states is given 
in Figure \ref{fig:4figures}.

\section{Example of historical evolution}
\vspace{-10mm}
\begin{figure}[htbp!]
\begin{center}
{\includegraphics[width=6cm]{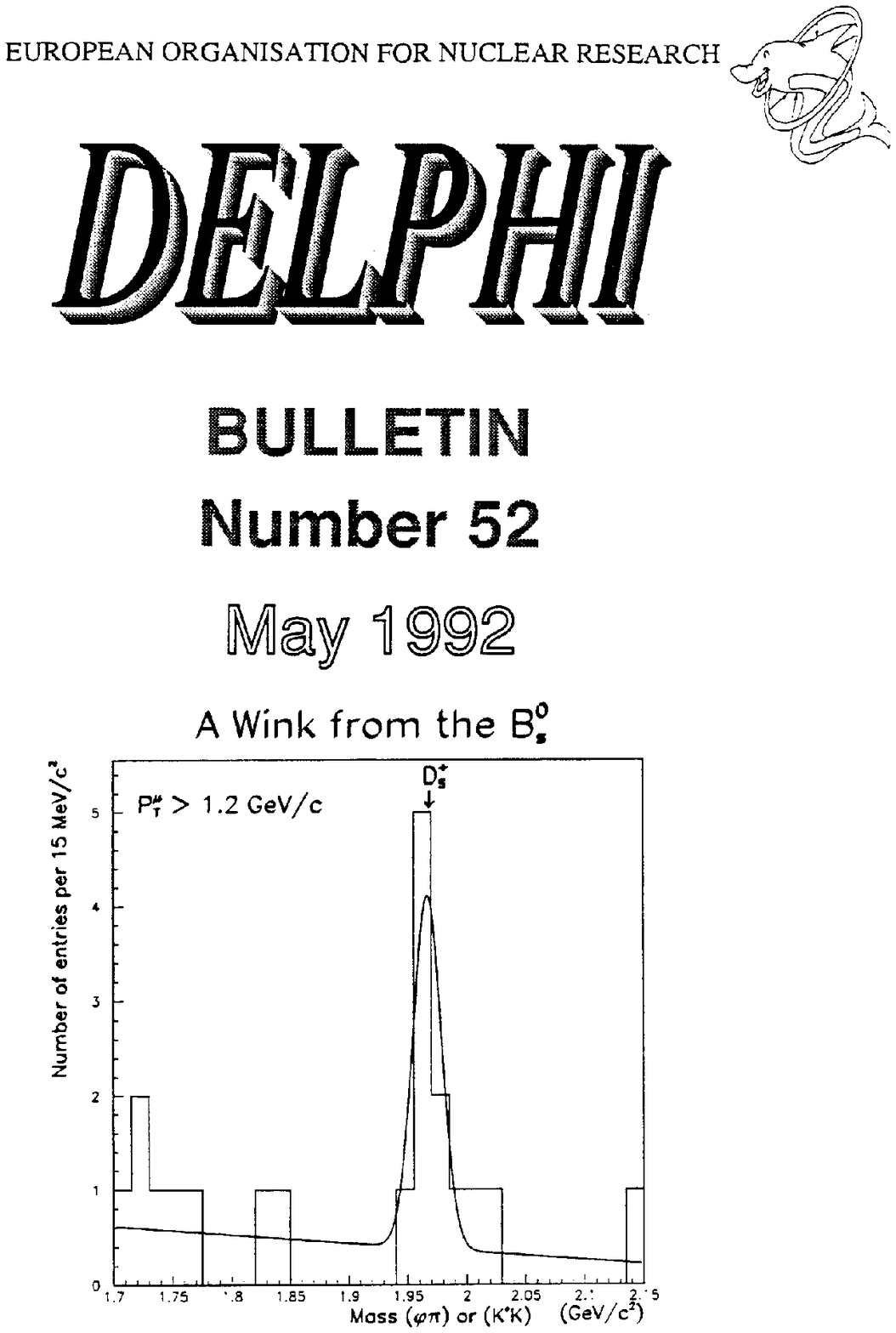}}
{\includegraphics[width=8cm]{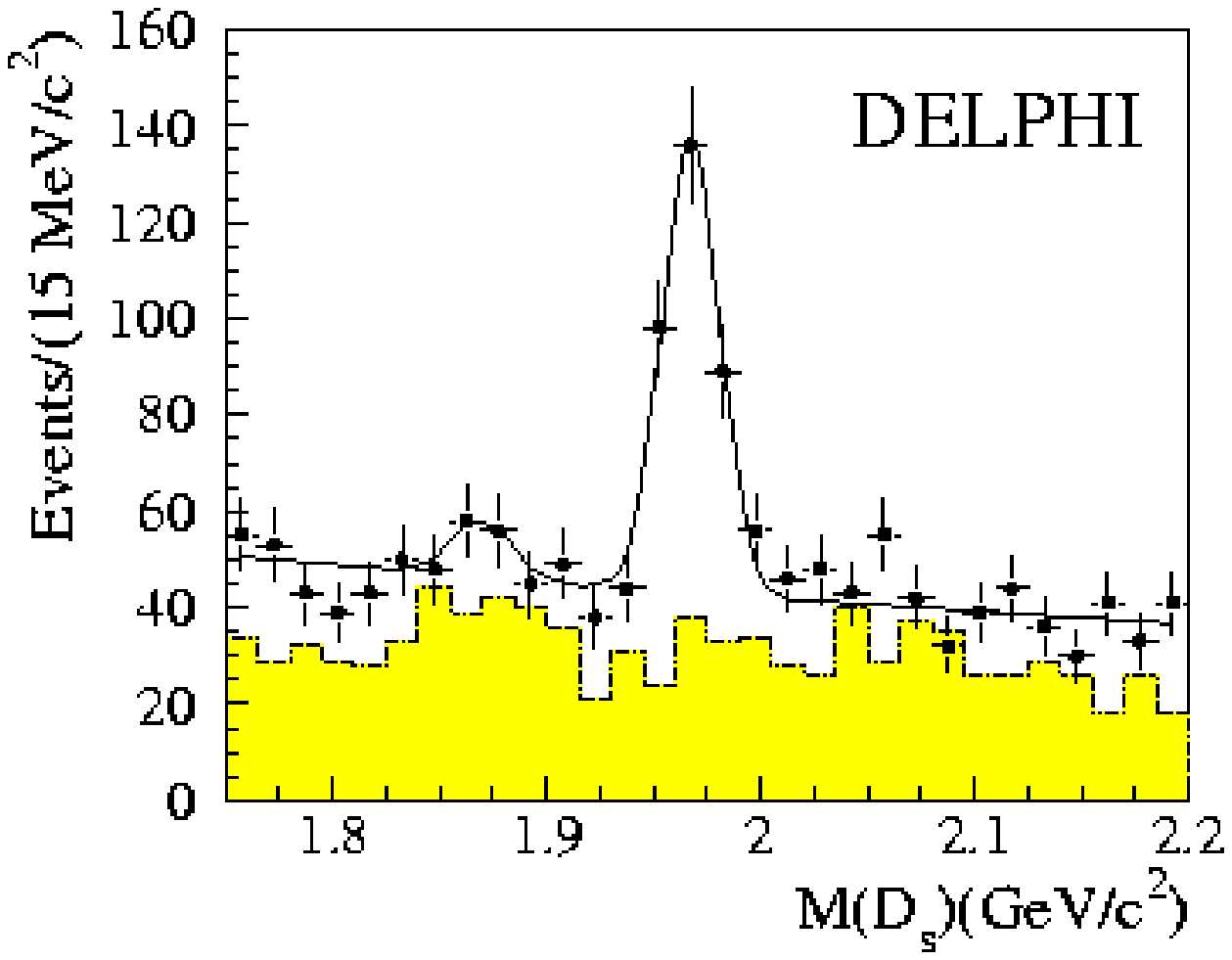}} 
\caption[]{\it{ The left plot shows the first signal of the $\bar{B}_s^0$ meson in 1992, seen in the 
semileptonic decay : $\bar{B}_s^0 \rightarrow D_s^+ \ell^- \overline{\nu_{\ell}}$, whereas 
the right plot shows the same signal few years later. }}
\end{center}
\end{figure}

\newpage
 We take the example of the ${B}_s^0$ meson to illustrate how our knowledge
 on the properties of B hadrons has evolved during the last ten years.
In 1992, 7 events $\bar{B}_s^0 \rightarrow D_s^+ \ell^- \overline{\nu_{\ell}}$,
constituted the first evidence for the $B_s^0$ meson. A few years later the same
 signal consists of more than 200 events.\\
 In the mean time our knowledge has much improved: the fraction of $B^0_s$ mesons in b jets is 
precisely measured as well as the $B^0_s$ mass and  lifetime.
\begin{itemize}
\item the $\bar{B}_s^0$ rate in $b$-jets amounts to: $f_s ~=~(9.7\pm 1.2)\%$, 
\item the $B_s^0$ meson mass is $m_{B_s^0}=(5369.6 \pm 2.4)$ MeV (CDF mainly)
\item the lifetime is $\tau(\bar{B}_s^0) = 1.464 \pm 0.057~ps$. 
\item the studies on $B_s^0-\bar{B}_s^0$ oscillations give $\Delta m_s > 15 ps^{-1} ~~~(95\%~\rm C.L.)$
\item the ratio $\Delta \Gamma_{B_s^0}/\Gamma_{B_s^0}<0.31~~~(95\%~{\rm C.L.})$
\end{itemize}

\section{Heavy hadron lifetimes}
\vspace{5mm}
Measurements of B lifetimes test the decay dynamics, giving 
important information on non-perturbative QCD corrections induced by 
the spectator quark (or diquark). Decay rates are expressed using the
OPE formalism, as a sum of operators developed in series of order 
$O(\Lambda_{QCD}/m_Q)^n$. In this formalism, no term of order $1/m_Q$ is 
present and spectator effects contribute at order 
$1/m_Q^3$ \footnote{Terms at order 1/$m_Q$ would appear if in this 
expansion the mass of the heavy hadron was used instead of the mass of the quark. 
The presence of such a term would violate the quark-hadron duality.}.
Non-perturbative operators are evaluated, most reliably, using 
lattice QCD calculations.

\subsection{Beauty hadron lifetimes}
Since the beginning of the LEP/SLD data taking an intense activity has been concentrated
on the studies of B hadron lifetimes. \\
Results are given in Table \ref{table:life} \cite{lifeWG}.

\begin{table}[htb!]
\caption{Summary of B hadron lifetime results (as calculated by the Lifetime 
Working Group \cite{lifeWG}).}
\begin{center}
\label{table:life}
\begin{tabular}{@{}ll}
 B Hadrons                          &    ~~~~~~ Lifetime~[ps]                 \\ \hline 
~~$\tau({B^0_d})$                     & 1.540 $\pm$ 0.014 ~ (0.9~$\%$) \\ 
~~$\tau({B^+})$                       & 1.656 $\pm$ 0.014 ~ (0.8~$\%$) \\
~~$\tau({B^0_s})$                     & 1.461 $\pm$ 0.057 ~ (3.9~$\%$) \\
~~$\tau({\Lambda^0_b})$               & 1.208 $\pm$ 0.051 ~ (4.2~$\%$) \\ \hline
 \multicolumn{2}{c}{$\tau({B^0_d})/\tau({B^+})$        ~~~~~=~ 1.073 $\pm$ 0.014}                 \\
\multicolumn{2}{c}{$\tau({B^0_d})/\tau({B^0_s})$       ~~~~~~=~ 0.949 $\pm$ 0.038 }                \\
\multicolumn{2}{c}{$\tau({\Lambda^0_b})/\tau({B^0_d})$ ~~~~~~=~ 0.798 $\pm$ 0.052 }                \\ 
\multicolumn{2}{c}{$\tau({b-bar})/\tau({B^0_d})$       ~=~ 0.784 $\pm$ 0.034 }                \\ \hline
\end{tabular}
\end{center}
\end{table}

\begin{figure}[htb!]
\begin{center}
\includegraphics[width=120mm]{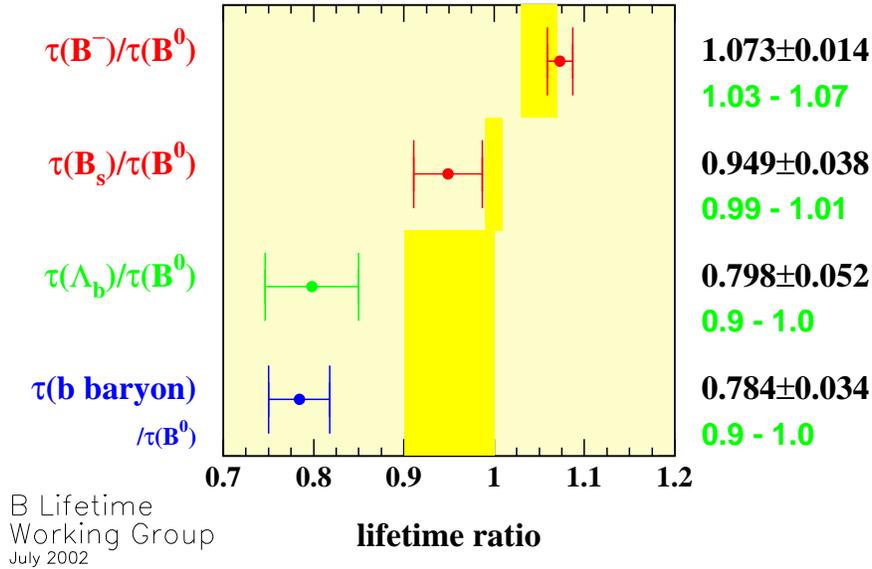}
\caption{\it {B hadrons lifetime ratios \cite{lifeWG}, compared with the theoretical predictions as given 
by the dark(yellow) bands.}}
\label{fig:liferatio}
\end{center}
\end{figure}

\begin{figure}[htbp!]
\begin{center}
\includegraphics[width=11cm]{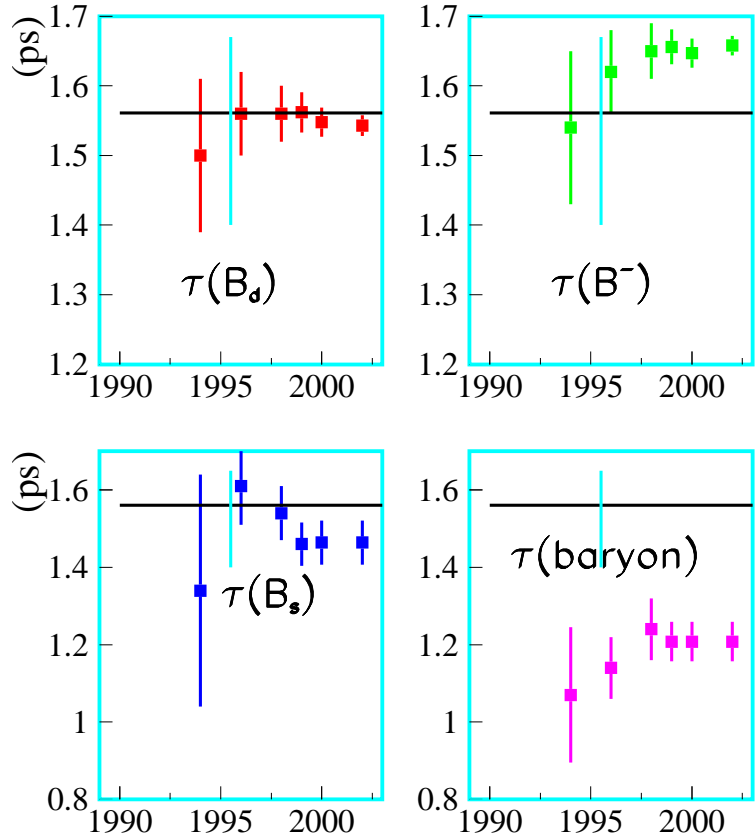}
\caption[]{\it{Evolution of the combined measurement of the different B hadron 
lifetimes over the years. The vertical band, in each plot, indicates the end of the data taking at LEP.}}
\label{fig:life_story}
\end{center}
\end{figure}

Figure \ref{fig:liferatio} gives the ratios of different B hadron lifetimes,
as compared with theory predictions (dark(yellow) bands).
\newpage
The achieved experimental precision is remarkable and LEP results are 
still dominating the scene. The fact that charged B mesons live longer than
neutral B mesons is now established at 5$\sigma$ level and is in agreement with theory.
The $B^0_d$ and $B^0_s$ lifetimes are expected (at $\simeq$1$\%$) and found (at $\simeq$4$\%$) 
to be equal. A significant measurement in which this ratio differs 
from unity will have major consequences for the theory. 
The lifetime of the b-baryons is measured to be shorter than the $B^0_d$ lifetime, but the size 
of this effect seems to be more important than predicted (2-3$\sigma$).
Recent calculations of high order terms give an evaluation of 
the b-baryon lifetime 
in better agreement with the experimental result\cite{vittorio}. \\
New results are expected from B-Factories (which could decrease the relative error on 
the lifetimes of the $B^0_d$ and $B^+$ to 0.4-0.5$\%$) and mainly from Tevatron (Run II) 
which could precisely measure all B hadron lifetimes, including those for the $\Xi_b$, 
$\Omega_b$ and the $B_c$.\\
In figure \ref{fig:life_story} the improvement on the precision of
the measured B hadron lifetimes over the years is shown.

\section{Determination of the CKM element: $|V_{cb}|$}
The $|V_{cb}|$ element of the CKM matrix can be accessed by studying the rates 
of inclusive and exclusive semileptonic $b$-decays.
\subsection{ $|V_{cb}|$ inclusive analyses.}
The first method to extract $|V_{cb}|$ makes use of the inclusive
semileptonic decays of B-hadrons and of the theoretical calculations
done in the framework of the OPE.
The inclusive semileptonic width $\Gamma_{s.l.}$ is expressed as:
\begin{eqnarray} 
 \Gamma_{s.l.} =  \frac{BR(b \rightarrow c l \nu)}{\tau_b}  = \gamma_{theory} |V_{cb}|^2 ; & \nonumber \\
  \gamma_{theory} = f(\alpha_s,m_b,\mu_{\pi}^2,1/m_b^3...).
\label{eq:vcbtheo}
\end{eqnarray} 
From the experimental point of view the semileptonic width has been measured by the LEP/SLD and 
$\Upsilon(4S)$ experiments with a relative precision of about 2$\%$ \cite{vcbWG}:
\begin{eqnarray}
   \Gamma_{sl} = & (0.431 \pm 0.008 \pm 0.007) 10^{-10} MeV  &  \small{\Upsilon(4S)}    \nonumber \\
   \Gamma_{sl} = & (0.439 \pm 0.010 \pm 0.007) 10^{-10} MeV  &  \small{\rm{LEP/SLD}}    \nonumber \\
   \Gamma_{sl} = & (0.434 \times (1 \pm 0.018)) 10^{-10} MeV  &  \small{\rm{ave.} }
\label{eq:gammsl}
\end{eqnarray}
The precision on the determination of $|V_{cb}|$ is mainly limited by theoretical 
uncertainties on the parameters entering in the expression of $\gamma_{theory}$ in equation \ref{eq:vcbtheo}.

\subsection{Moments analyses}
\begin{figure}[htb!]
\begin{center}
\includegraphics[width=95mm]{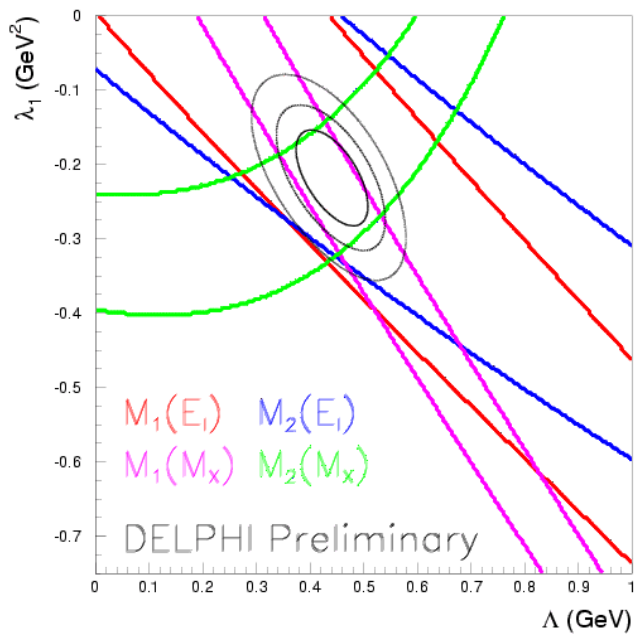} 
\caption{\it {Constraints in the $\bar{\Lambda}-\lambda_1$ plane obtained: 
by the DELPHI Coll. using the measured values of the first two moments of 
the hadronic mass and lepton energy spectra. The bands represent 
the 1$\sigma$ regions selected by each moment and the ellipses show the 
39$\%$, 68$\%$ and 90$\%$ probability regions of the global fit.}}
\label{fig:moments}
\end{center}
\end{figure}

Moments of the hadronic mass spectrum, of the lepton energy spectrum and of 
the photon energy in the $b \rightarrow s \gamma$ decay are sensitive to the non
perturbative QCD parameters contained in the factor $\gamma_{theory}$ of 
equation \ref{eq:vcbtheo} and in particular to the mass of the $b$ and $c$ quarks 
and to the Fermi motion of the heavy quark inside the hadron, $\mu_{\pi}^2$ 
\footnote{In another formalism,
based on pole quark masses, the $\bar{\Lambda}$ and $\lambda_1$ parameters are used, which can
be related to the difference between hadron and quark masses and to $\mu_{\pi}^2$, respectively.}.\\
Results from DELPHI collaboration are shown in Figure \ref{fig:moments}.\\
Similar results (and with comparable precision) have been obtained by CLEO (which did a 
pioneering work in this field) and by the BaBar Coll. \cite{ref:stocchi}.

\noindent
Using the experimental results on $\bar{\Lambda}$ and $\lambda_1$:
\begin{equation}
 |V_{cb}| = (40.7 \pm 0.6 \pm 0.8(\rm{theo.})) 10^{-3} \rm{(inclusive)}
\label{eq:vcbinclres}
\end{equation}
This result corresponds to an important improvement on the determination of the $|V_{cb}|$ element. Part of the
theoretical errors (from $m_b$ and $\mu_{\pi}^2$) is now absorbed in the experimental error and the theoretical error is 
reduced by a factor two. The remaining theoretical error could be further reduced if the 
parameters controlling the $1/m_b^3$ corrections are extracted directly from experimental data.

\subsection{$|V_{cb}|$: $B \rightarrow D^* \ell \nu$ analyses.}
An alternative method to determine $|V_{cb}|$ is based on exclusive $\overline{ B^0_d} \rightarrow D^{*+} \ell^- 
\overline{\nu_l}$ decays. Using HQET, an expression for the differential decay rate can be derived
\begin{equation}
\frac{d\Gamma}{dw} = \frac{G_F^2}{48 \pi^2} |V_{cb}|^2 |F(w)|^2 G(w) ~;~ w = v_B.v_D 
\label{eq:exclvcb}
\end{equation}
\begin{figure}[htb]
\begin{center}
\includegraphics[width=100mm]{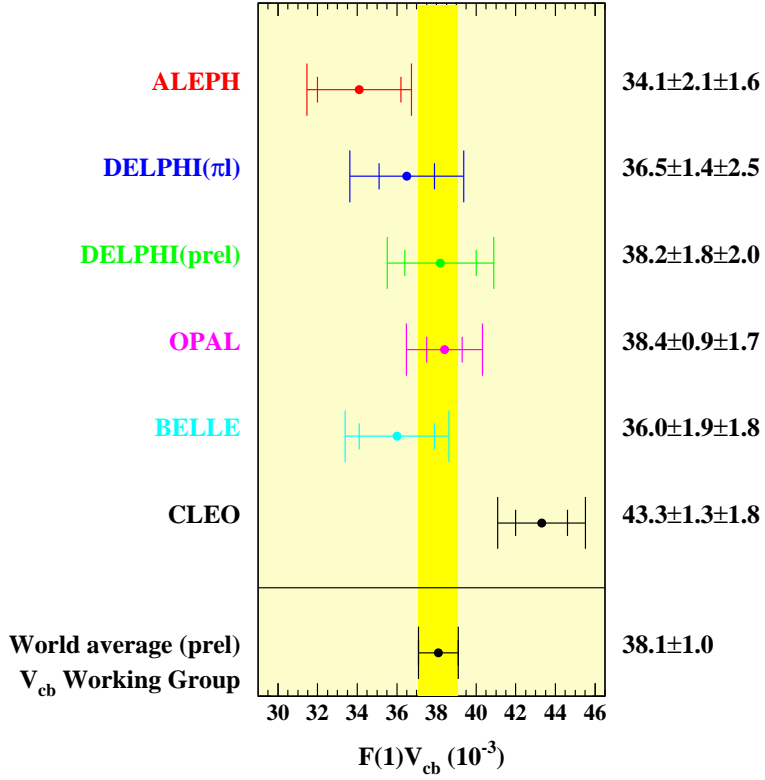}
\end{center}
\caption{\it  {Summary of the measurements of $F(1) \times |V_{cb}|$ \cite{vcbWG}.}}
\label{fig:vcb_ave}
\end{figure}
$w$ is the relative velocity between the B ($v_B$) and the D systems ($v_D$). G($w$) is a kinematical
factor and F($w$) is the form factor describing the transition. At zero recoil ($w$=1)
F(1) goes to unity. The strategy is then to measure $d\Gamma/dw$, to extrapolate 
at zero recoil and to determine $F(1) \times |V_{cb}|$.\\
The experimental results are summarised in Figure \ref{fig:vcb_ave}.
Using F(1) = 0.91 $\pm$ 0.04 \cite{latticeF1}, it gives \cite{vcbWG}:
\begin{equation}
 |V_{cb}| = (41.9 \pm 1.1 \pm 1.9(F(1)) 10^{-3}  \rm{(exclusive)}
\label{eq:vcbexclres}
\end{equation}
Combining the two determinations of $|V_{cb}|$ (a possible correlation 
between the two determinations has been neglected) it gives \cite{ref:stocchi}:
\begin{eqnarray}
 |V_{cb}| = (40.9 \pm 0.8) 10^{-3}   \tiny{\rm{(exclusive+inclusive)}}
\label{eq:vcbave}
\end{eqnarray}

\section{Measurement of $|V_{ub}|$.}
\label{sec:vub}

\begin{figure}[htbp!]
\begin{center}
\includegraphics[width=100mm]{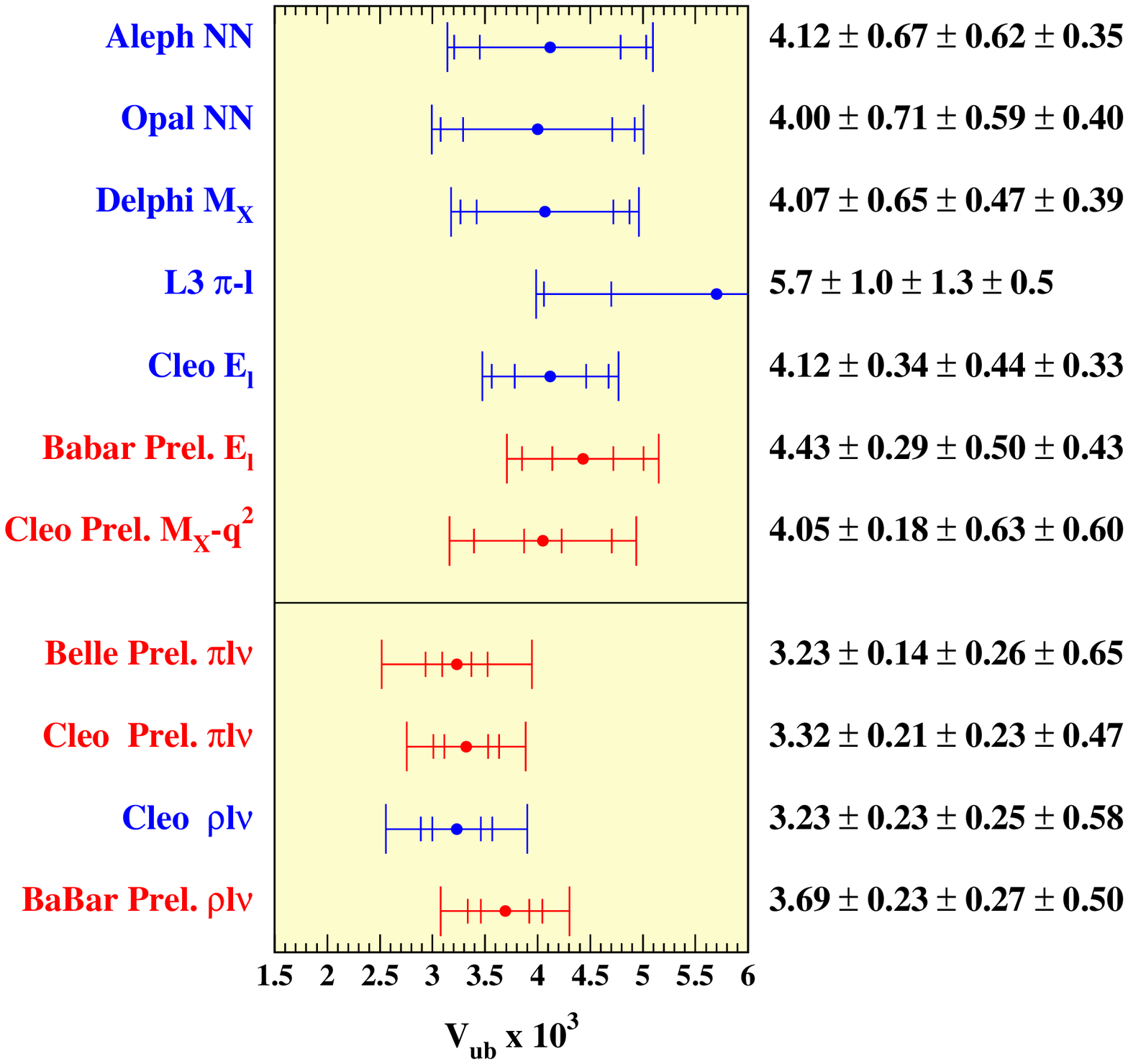}
\caption{\it{Summary of $|V_{ub}|$ measurements \cite{marco}.}}
\label{fig:vub_ave}
\end{center}
\end{figure}

The CKM matrix element $|V_{ub}|$ has been measured at LEP using 
semileptonic $b$ to $u$ decays. This measurement is rather
difficult because one has to suppress the large background
from the more abundant semileptonic $b$ to $c$ quark transitions. By using kinematical 
and topological variables, the LEP experiments have succeeded 
in measuring the semileptonic $b$ to $u$ branching ratio \cite{ref:ckmworkshop}, and obtain : 
\begin{center} 
 $ BR(b\rightarrow l^-\bar{\nu} X_u) = (1.71 \pm 0.53 ) \:10^{-3} $
\end{center} 
 
Using models based on the Operator Product Expansion, a value for $|V_{ub}|$ is obtained  :
\begin{eqnarray}
 |V_{ub}| &=& ( 40.9 \pm 6.1 \pm 3.1 ) \times 10^{-4}  \quad \mbox{LEP} \ .
\end{eqnarray}
Prior to this analysis, the $V_{ub}$ matrix element was firstly obtained, by CLEO and 
ARGUS collaborations, by looking at the spectrum of the lepton in B semileptonic decays. 
The difference between D meson and $\pi$ masses is reflected in the momentum of 
the lepton from the B decays. This analysis has been recently revised by the CLEO Coll..
An alternative method to determine $|V_{ub}|$ consists in the reconstruction of the charmless 
semileptonic B decays: $B \rightarrow \pi (\rho) \ell \nu$. This analysis has been performed 
by the CLEO Coll. and now by the b-factories. \\
Figure \ref{fig:vub_ave} shows the full set of results on $V_{ub}$ \cite{marco}. 

\section{Study of $B^0-\overline{B^0}$ oscillations}
\label{sec:oscillations}

The probability that a $B^0$ meson oscillates into a $\overline{B}^0$ or
remains a ${B}^0$ is given by:
\begin{equation}
P_{{B}^0_q \rightarrow {B}^0_q(\overline{{B}}^0_q)} =
\frac{1}{2}e^{-t/\tau_q} (1 \pm \cos \Delta {m}_q t)
\label{eq:oscillation}
\end{equation}
Where $t$ is the proper time, $\tau_q$ the lifetime of the 
${B}^0_q$ meson, and $\Delta {m}_q = {m}_{B^0_1}- m_{{B}^0_2}$ the mass difference 
between the two physical mass eigenstates
\footnote{$\Delta {m}_q$ is usually given in ps$^{-1}$: 1 ps$^{-1}$ corresponds to 
6.58 10$^{-4}$eV.}.
To derive this formula the effects of CP violation and lifetime differences
for the two states have been neglected. \\
Integrating expression \ref{eq:oscillation}, over the decay time, 
the probability to observe a $\bar{B}^0_{d(s)}$ meson starting from a $B^0_{d(s)}$ 
meson is given by $\chi_d(s) = x^2_{d(s)}/(2+x^2_{d(s)})$, where $x_{d(s)} = 
\Delta m_{d(s)} \tau(B_{d(s)}^0)$. At Z energies, both $B_d^0$ and $B^0_s$ 
mesons are produced  with fractions $f_{B_d}$ and $f_{B_s}$. The average mixing parameter   
$\chi$ is defined as : $\chi = f_{B_d} \chi_d +  f_{B_s}  \chi_s$.
It has to be noted that for fast $B^0_s$ oscillations $\chi_s$ takes values close 
to 0.5 and $\chi_s$ becomes very insensitive to $x_s$. Even a very precise measurement 
of $\chi_s$ does not allow a determination of $\Delta m_s$. \\
It is then clear that only the time evolution of the $B^0-\overline{B}^0$
oscillations allow to measure  $\Delta m_d$ and $\Delta m_s$.\\
A time dependent study of $B ^0 - \overline{B}^0$ oscillations requires: 
\begin{itemize}
\item the measurement of the proper time t, 
\item to know if a $B ^0$ or a  $\overline{B}^0$ decays at time t  (decay tag)
\item to know if a $b$ or a $\overline{b}$ quark has been produced at t = 0 (production tag). 
\end{itemize}

 In the Standard Model,  $B^0-\bar{B}^0$ oscillations 
occur through a second-order process - a box diagram - with a loop of W and up-type quarks. 
 The box diagram with the exchange of a $top$ quark gives the dominant contribution :
\begin{eqnarray}
\Delta {m}_d & ~\propto  ~V_{td}^2 f^2_{{B}_d} {B}_{B_d}  
        ~\propto  ~  V_{cb}^2 \lambda^2 [(1 - \bar{\rho})^2 + \bar{\eta}^2]f^2_{{B}_d} {B}_{B_d} &  \nonumber \\
\Delta {m}_s &~\propto  ~V_{ts}^2 f^2_{{B}_s} {B}_{B_s}  
              ~\propto  ~  V_{cb}^2 f^2_{{B}_s} {B}_{B_s}  &\nonumber \\
\frac{\Delta m_d}{\Delta m_s} &~\propto 1/\xi^2|\frac{V_{td}}{V_{ts}}|^2 ~\propto  ~ 
1/\xi^2 \lambda^2 [(1 - \bar{\rho})^2 + \bar{\eta}^2] &
\label{eq:dmddmsxi}
\end{eqnarray}

where  $\xi=\frac{ f_{B_s}\sqrt{B_{B_s}}}{ f_{B_d}\sqrt{B_{B_d}}}$ .

Thus, the measurement of $\Delta m_d$ and $\Delta m_s$ gives access to 
the CKM matrix elements $|V_{td}|$
and $|V_{ts}|$ respectively.
The difference in the $\lambda$ dependence of these expressions ($\lambda \sim 0.22$) implies that 
$\Delta {m}_s \sim 20 ~\Delta {m}_d$. It is then clear that a very good proper time resolution is needed 
to measure the $\Delta {m}_s$ parameter.
On the other hand the measurement of the ratio $\Delta m_d/\Delta m_s$ gives the same 
constraint as $\Delta m_d$ but this ratio is expected to have smaller theoretical uncertainties 
since the ratio $\xi$ is better known than the absolute value of $f_B \sqrt B_B$.

\subsection*{$\Delta {m}_d$ measurements}

\begin{figure}[htbp!]
\begin{center}
\epsfig{file=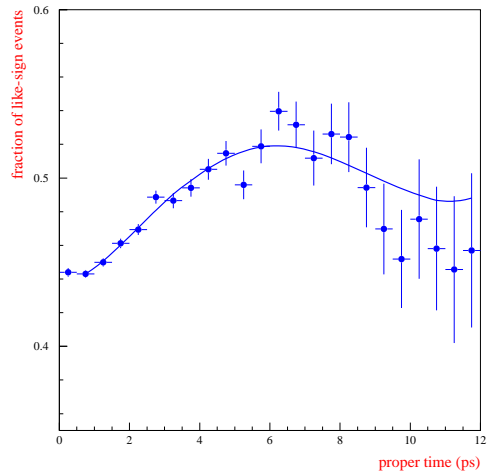,width=7cm}
\caption{\it {This plot shows the fraction of like-sign events as a function of the
proper decay time. Points with error bars are the data. The curve corresponds to 
the result of the fit to $\Delta {m}_d$.}}
\label{fig:dmd}
\end{center}
\end{figure}

\begin{figure}[htbp!]
\begin{center}
\epsfig{file=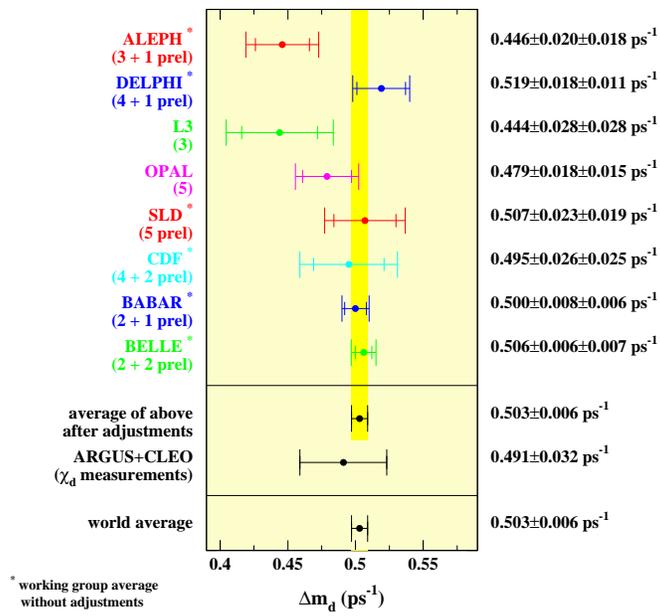,width=9cm}
\caption{\it {Summary of the $\Delta {m}_d$ results from LEP, SLD, CDF, BABAR and BELLE.}}
\label{fig:dmdsummary}
\end{center}
\end{figure}

Analyses using different events sample have been performed at LEP. 
A typical time distribution is shown in Figure \ref{fig:dmd}.  
$B^0_d - \overline{B}^0_d$ oscillations with a frequency 
$\Delta {m}_d$ are clearly visible. This can be a textbook plot ! 
The present summary of these results on $\Delta {m}_d$, is shown in Figure \ref{fig:dmdsummary}. 
Combining LEP, CDF and SLD measurements it follows that \cite{osciWG}:
\begin{equation}
\Delta m_d = (0.498 \pm 0.013) \:ps^{-1}
\end{equation}
\begin{figure}[htbp!]
\begin{center}

$\Delta {m}_d$ has been first measured with high precision by the LEP/SLD/CDF experiments. 
The new and precise measurements performed at the B-Factories confirmed these measurements 
and improved the precision by a factor two. 
the combined result is now : $\Delta m_d = (0.503 \pm 0.006) \: ps^{-1}$.
The evolution, over the years, of the combined $\Delta m_d$ frequency measurement 
is shown in Figure \ref{fig:dmd_story}.

\includegraphics[width=8cm]{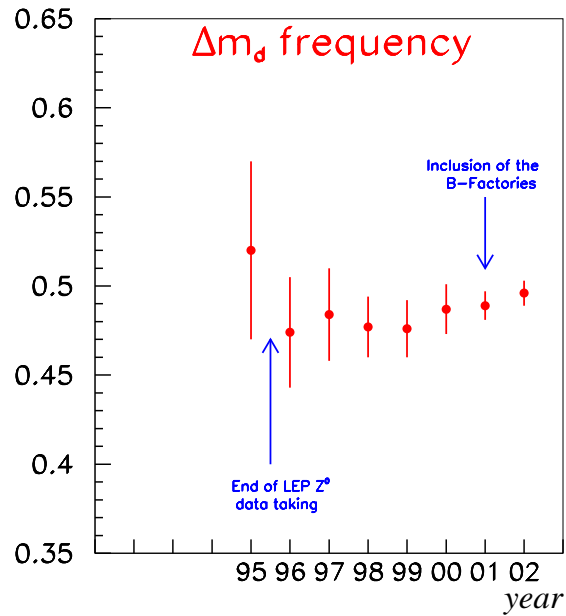}
\caption[]{\it{The evolution of the combined $\Delta m_d$ frequency measurement over the years.}}
\label{fig:dmd_story}
\end{center}
\end{figure}

\subsection*{Analyses on $\Delta {m}_s$}

The search for $B^0_s-\overline{B^0_s}$  oscillations is more difficult because 
the oscillation frequency is much higher. In the Standard Model 
one expects $\Delta {m}_s \sim 20 ~\Delta {m}_d$. The proper time resolution will
therefore play an essential role. Five different types of analyses have been performed 
at LEP/SLD. An overview is given in Table \ref{tab1}.

\begin{table}[htb!]
\begin{center}
\begin{tabular}{cccccc} \hline
Analysis & N(events) & $P({B}_S)$ & $\varepsilon_1$ & $\varepsilon_2$ & $\sigma_t (t~<~1 \rm{ps})$ \\ \hline
Dipole           & $\sim 700000$ &  $\sim 10\%$ & $\sim 70\%$ & $\sim 60\%$  & $\sim 0.25$ ps \\ 
Inclusive lepton & $\sim 50000$ & $\sim 10\%$ & $\sim 70\%$ & $\sim 90\%$ & $\sim 0.25$ ps\\
${D}^\pm_s h^\mp$ & $\sim3000$ & $\sim 15\%$ & $\sim 72\%$ & $\sim 90\%$ & $\sim 0.22$ ps\\
${D}^\pm_s \ell^\mp$ & $\sim 400$ & $\sim 60 \%$ & $\sim 78 \%$ & $\sim 90 \%$ & $\sim 0.18$ ps\\
Exclusive ${B}^0_S$ & $\sim 25$ & $\sim 70 \%$ & $\sim 78 \%$ & $\sim 100 \%$ & $\sim 0.08$ ps\\
\hline
\end{tabular}
\caption{\it {Characteristics of the different analyses are given in terms of statistics (N), 
${B}^0_s$ purity [$(P({B}_s)$] , tagging purities - i.e. the fraction of correctly 
tagged events - at the production and decay time $(\varepsilon_1, \varepsilon_2)$ and 
average time resolution within the first picosecond.}}
\label{tab1}
\end{center}
\end{table}

The so-called amplitude method \cite{ref:amp} has been developed to combine data from
different experiments. It corresponds to the following change in 
equation \ref{eq:oscillation}: $$1 \pm \cos \Delta {m}_s t \rightarrow 1 \pm A \cos \Delta {m}_s t$$
A and $\sigma_A$ are measured at fixed values of $\Delta {m}_s$.
In case of a clear oscillation signal, the measured 
amplitude is compatible with A = 1 at the corresponding value of $\Delta m_s$.
With this method it is also easy to set an exclusion limit. 
The values of $\Delta {m}_s$ excluded at 95\% C.L. are those satisfying the condition 
A($\Delta{m}_s$) + 1.645 $\sigma_A (\Delta {m}_s) < 1$. 
Furthermore, the sensitivity of the experiment can be defined as the value of
$\Delta {m}_s$ corresponding to 1.645 $\sigma_A (\Delta {m}_s) = 1$ (for
A($\Delta {m}_s) = 0$, namely supposing that the ``true'' value of
$\Delta {m}_s$ is well above the measurable value of $\Delta {m}_s$). 
\begin{figure}[htb!]
\begin{center}
\includegraphics[width=11cm]{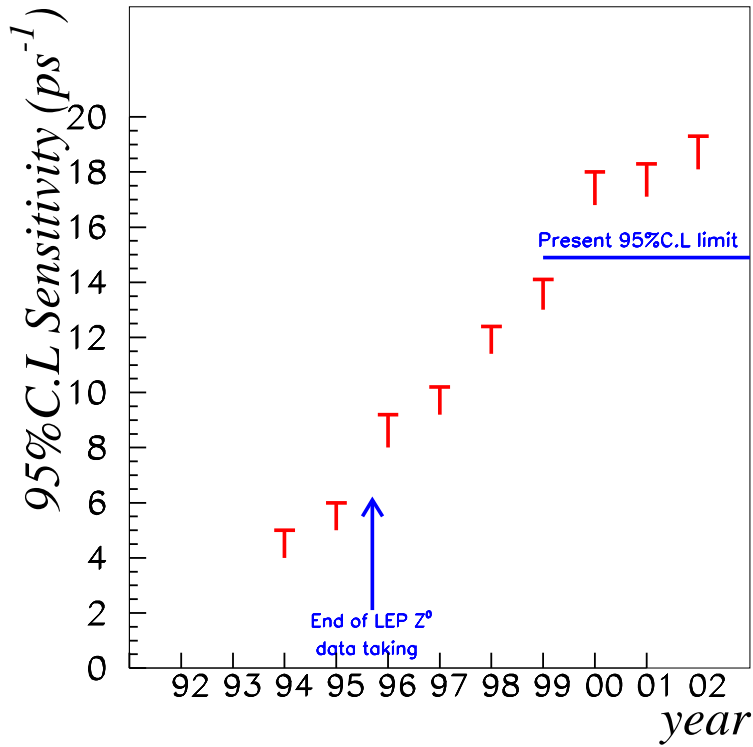}
\caption[]{\it{The evolution, over the years, of the combined $\Delta m_s$ sensitivity.}}
\label{fig:dms_story}
\end{center}
\end{figure}

\begin{figure}[htb!]
\begin{center}
\epsfig{file=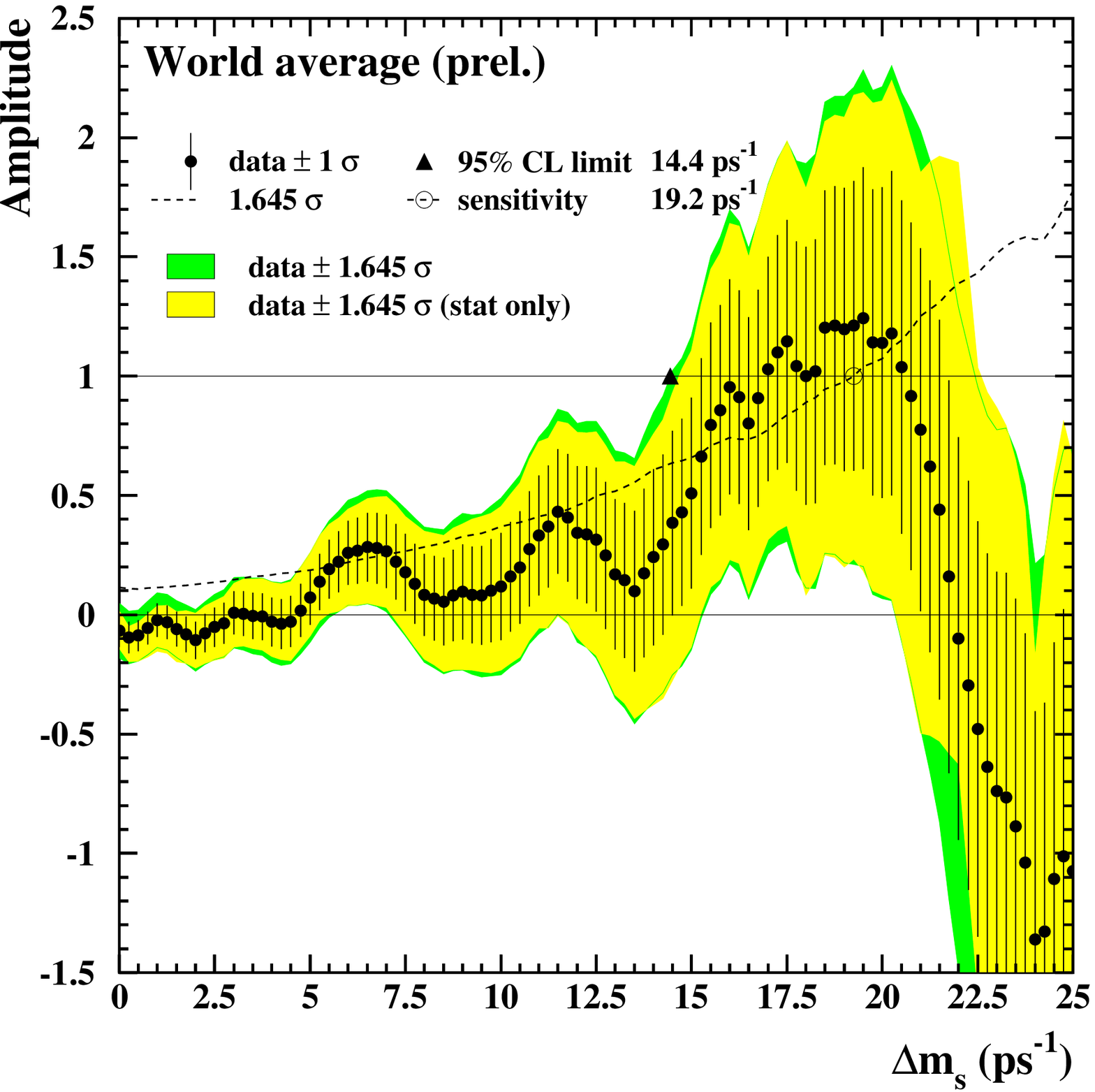,height=12cm} 
\caption{\it{The plot shows the combined $\Delta {m}_s$ results from 
LEP/SLD/CDF analyses displayed as an amplitude versus $\Delta {m}_s$ plot. 
Points with error bars are the data; the lines show the 95\% C.L. curves (darker regions include 
systematics) \cite{osciWG}. The dotted curve shows the sensitivity.}}
\label{fig:dms}
\end{center}
\end{figure}

\begin{figure}[htb!]
\begin{center}
\epsfig{file=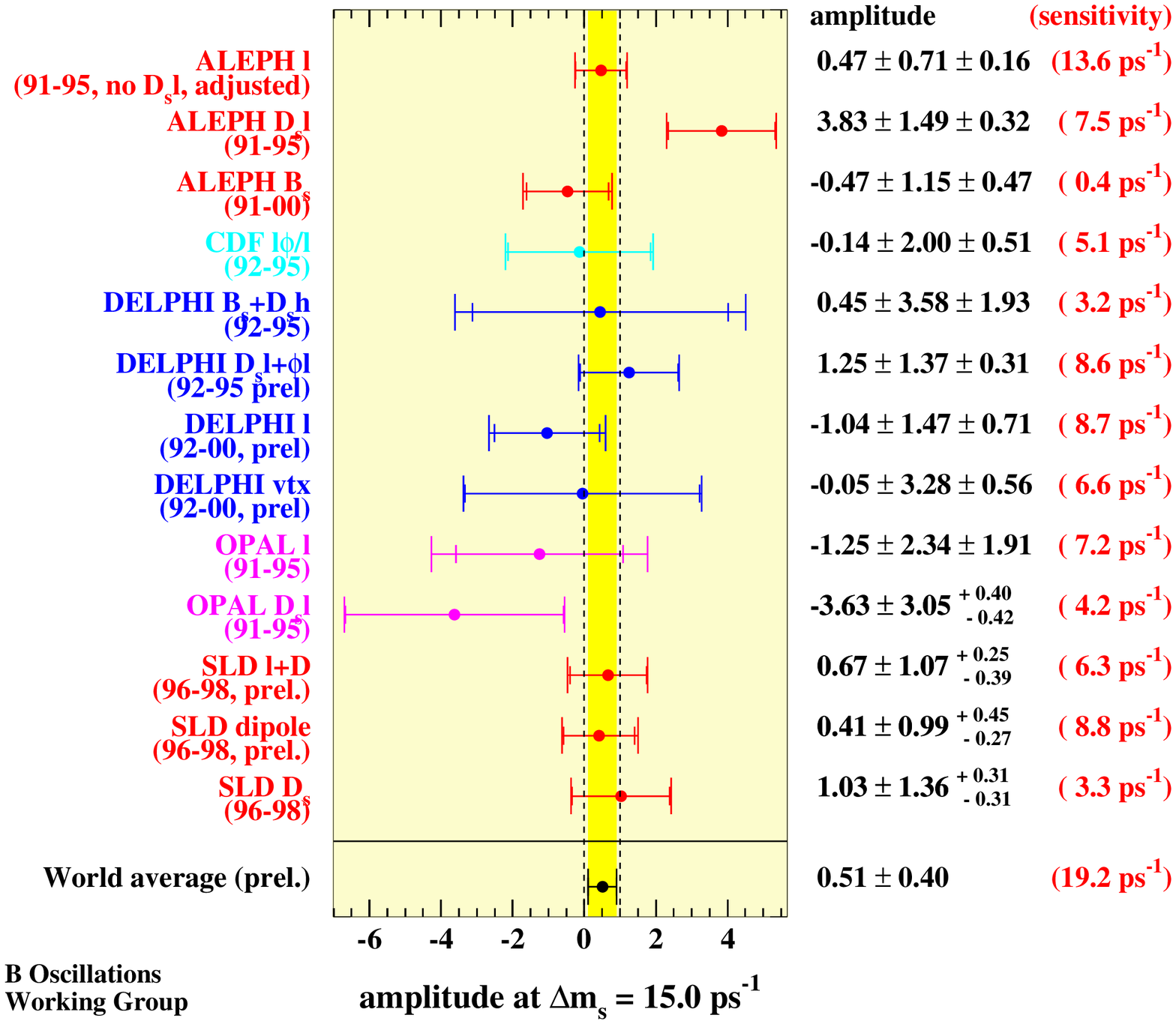,height=10cm}
\caption{\it{This plot shows the summary of results on the $\Delta {m}_s$,
per experiment. Errors are given at $\Delta {m}_s = 15~{ps}^{-1}$ 
(the sensitivity is also indicated). }}
\label{fig:dmssummary}
\end{center}
\end{figure}

During the last seven years impressive improvements in the analysis techniques 
allowed to improve the sensitivity of this search, as it can be seen in
Figure \ref{fig:dms_story}.\\
The combined result of the LEP/SLD/CDF \cite{osciWG} analyses, displayed as an 
amplitude vs $\Delta m_s$ plot, is shown in Figure \ref{fig:dms} and is:
$$\Delta {m}_s > 14.4~\rm{ps}^{-1}~~\rm{at}~~95\%~~\rm{C.L.}$$
The sensitivity is at $19.2~\rm{ps}^{-1}$. \\
The summary of the present results on $\Delta m_s$ is shown in Figure \ref{fig:dmssummary}.

The present combined limit implies that $B_s^0$ oscillates at least 30 
times faster than $B_d^0$ mesons.\\
The significance of the ``signal'', appearing around 17 ps$^{-1}$, is about 2.5 $\sigma$ and
no claim can be made on the observation of $B^0_s-\bar{B^0_s}$ oscillations.\\
Tevatron experiments, are thus expected to measure soon $B^0_s-\bar{B^0_s}$ oscillations...

\section{The CKM Matrix} 
\label{sec:ickm} 
In the Standard Model, the weak interactions among quarks are encoded
in a 3 $\times$ 3 unitary matrix: the CKM matrix.\\
\newpage
\noindent
 The existence of 
this matrix conveys the fact that quarks weak interaction eigenstates are a 
linear combination of their mass eigenstates \cite{Cabibbo,km}. \\
\begin{equation}
V_{CKM} =
\left ( \begin{array}{ccc}
V_{ud} ~~ V_{us} ~~ V_{ub} \\
V_{cd} ~~ V_{cs} ~~ V_{cb} \\
V_{td} ~~ V_{ts} ~~ V_{tb}
\end{array} \right )
\end{equation}

\vspace{5mm}
The CKM matrix can be parametrized in terms of four free parameters. 
Here, the improved Wolfenstein \cite{ref:Wolf} parametrization, expressed in terms of 
the four parameters $\lambda$, $A$, $\rho$ and $\eta$ (which accounts for the CP violating 
phase) , will be used:

\begin{equation}
\begin{array}{cccc}
\hspace{-15mm}
V_{CKM} =
&
\left ( \begin{array}{cccc}
~~~~1 - \frac{\lambda^{2}}{2} - \frac{\lambda^4}{8} ~~~~~~~~~~~~~~~~~~~~~~~~~~~~~~  \lambda ~~~~~~~~~~~~~~~~~~~~~~~~     
A \lambda^{3} (\rho - i \eta) \\
   - \lambda +\frac{A^2 \lambda^5}{2}(1-2 \rho) -i A^2 \lambda^5 \eta~~~~~~~~~~  1 - \frac{\lambda^{2}}{2}
   -\lambda^4(\frac{1}{8}+\frac{A^2}{2})
 ~~~~~~~~           A \lambda^{2}       \\
A \lambda^{3} [1 - (1-\frac{\lambda^2}{2})(\rho +i \eta)] ~~~ -A \lambda^{2}(1-\frac{\lambda^2}{2})(1 + \lambda^{2}(\rho +i \eta))
 ~~~    1-\frac{A^2 \lambda^4}{2}
\end{array} \right )
& 
+ O(\lambda^{6}).
\end{array}
\label{eq:eq8}
\end{equation}

The CKM matrix elements can be expressed as: 
\begin{equation}
  V_{us}~=~ \lambda ~ 
  V_{cb}~=~~A \lambda^2,~
  V_{ub}~=~A \lambda^3(\overline{\rho}-i \overline{\eta})/(1-\lambda^2/2),~
  V_{td}~=~A \lambda^3(1-\overline{\rho}+i \overline{\eta})
\end{equation}
where the parameters $\overline{\rho}$ and $\overline{\eta}$ have been introduced \cite{ref:blo}
\footnote{ $ \overline{\rho} = \rho ( 1-\frac{\lambda^2}{2} ) ~~~;~~~ \overline{\eta} = 
\eta ( 1-\frac{\lambda^2}{2} ).$ }.

 The parameter $\lambda$ is precisely determined to be $0.2210 \pm 0.0020$
\footnote{due to the disagreement between the different determinations 
$\lambda$ has been recently evaluated to be \cite{ref:ckmworkshop}: $0.2237 \pm 0.0033$}
 using semileptonic kaon decays. The other parameters: $A$, $\overline{\rho}$ 
and $\overline{\eta}$ were rather unprecisely known.\\
The Standard Model predicts relations between the different processes 
which depend upon these parameters; CP violation is accommodated in the CKM
matrix and its existence is related to $\bar{\eta} \neq 0$.
The unitarity of the CKM matrix can be visualized as a triangle in the
$\bar{\rho}-\bar{\eta}$ plane. Several quantities, depending upon $\bar{\rho}$
and $\bar{\eta}$ can be measured and they must define compatible values for
the two parameters, if the Standard Model is the correct description of these
phenomena. Extensions of the Standard Model can provide different predictions
for the position of the upper vertex of the triangle, given by the 
$\bar{\rho}$ and $\bar{\eta}$ coordinates.
\begin{figure}[htb!]
\begin{center}
\includegraphics[width=120mm]{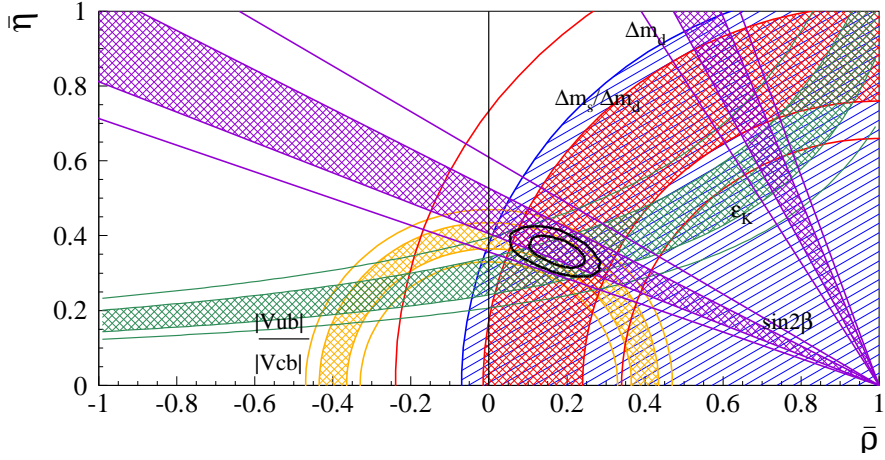}
\end{center}
\caption{\it{The allowed regions for $\overline{\rho}$ and $\overline{\eta}$
(contours at 68\%, 95\%) are compared with the uncertainty bands 
for $\left | V_{ub} \right |/\left | V_{cb} \right |$, 
$\epsilon_K$, $\Delta {m_d}$,the limit on $\Delta {m_s}/\Delta {m_d} $ and sin2$\beta$.}}
\label{fig:bande}
\end{figure}

\begin{figure}[htb!]
\begin{center}
\includegraphics[width=120mm]{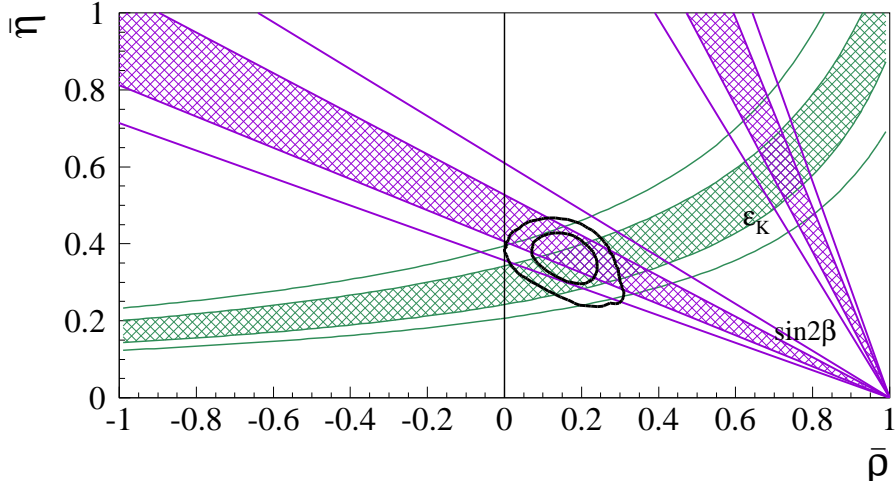}
\end{center}
\caption{\it {The allowed regions for $\overline{\rho}$ and $\overline{\eta}$
(contours at 68\%, 95\%) as selected by the measurement of $\left | V_{ub} \right |/\left | V_{cb} \right |$, 
$\Delta {m_d}$, the limit on $\Delta {m_s}/\Delta {m_d} $ are compared with the bands (at 1 and 2$\sigma$) 
selected from CP violation in the kaon ($\epsilon_K$) and in the B (sin2$\beta$) sectors.}}
\label{fig:testcp}
\end{figure}

\begin{table*}[htb]
\newcommand{\m}{\hphantom{$-$}}
\newcommand{\cc}[1]{\multicolumn{1}{c}{#1}}
\begin{center}
\begin{tabular}{@{}lllll}
\hline
 Parameter &  Value & Gaussian &  Uniform      & Ref. \\
           &        & $\sigma$ &  half-width   &      \\
\hline
    $\lambda$               & $0.2210$   &  0.0020    &         -     &  \cite{ref:ckmworkshop} \\
\hline
$\left | V_{cb} \right |$(excl.) & $ 42.1  \times 10^{-3}$  & $ 2.1 \times 10^{-3}$ 
                                 &              -           & \cite{ref:ArtusoBarberio}\\
$\left | V_{cb} \right |$(incl.) & $ 40.4  \times 10^{-3}$  & $ 0.7 \times 10^{-3}$ 
                                 & $ 0.8 \times 10^{-3}$    & \cite{ref:ArtusoBarberio}\\
 \hline
$\left | V_{ub} \right |$(excl.) & $ 32.5  \times 10^{-4}$ & $ 2.9 \times 10^{-4}$ 
                                 & $ 5.5 \times 10^{-4}$ & \cite{ref:ckmworkshop}\\
$\left | V_{ub} \right |$(incl.) & $ 40.9  \times 10^{-4}$ & $ 4.6 \times 10^{-4}$ 
                                 & $ 3.6 \times 10^{-4}$ & \cite{ref:ckmworkshop}\\
 \hline
$\Delta m_d$                      & $0.503~\mbox{ps}^{-1}$ & $0.006~\mbox{ps}^{-1}$ 
                                  & -- & \cite{osciWG}  \\
$\Delta m_s$  & $>$ 14.4 ps$^{-1}$ at 95\% C.L. & \multicolumn{2}{c}
{sensitivity 19.2 ps$^{-1}$} & \cite{osciWG}  \\
$m_t$ & $167~GeV$ & $ 5~GeV$ & -- & \cite{ref:top} \\
$f_{B_d} \sqrt{\hat B_{B_d}}$ & $235~MeV$  & $33~MeV$ &  $^{+0}_{-24}~MeV$  & \cite{ref:lellouch} \\
$\xi=\frac{ f_{B_s}\sqrt{\hat B_{B_s}}}{ f_{B_d}\sqrt{\hat B_{B_d}}}$ 
                                  & 1.18   & 0.04 & $^{+0.12}_{-0.00}$ & \cite{ref:lellouch} \\
 \hline
$\hat B_K$                    & 0.86   & 0.06 & 0.14 & \cite{ref:lellouch} \\
 \hline
         sin 2$\beta$             & 0.734   & 0.054 & - & \cite{ref:sin2b} \\ \hline
\hline
\end{tabular} 
\caption {\it{ Values of the relevant quantities used in the fit of the CKM parameters.
In the third and fourth columns the Gaussian and the flat parts of the uncertainty are given, 
respectively \cite{parodi}. The values and the errors on $V_{cb}$ are taken from \cite{ref:ArtusoBarberio}
and are slightly different with respect to those given in equations 
\ref{eq:vcbinclres},\ref{eq:vcbexclres}.}} 
\label{tab:inputs} 
\end{center}
\end{table*}
Different constraints can be used to select the allowed region for the apex of the triangle 
in the $\bar{\rho}$-$\bar{\eta}$ plane. 
Five have been used so far: $\epsilon_k$, $|V_{ub}|/|V_{cb}|$, $\Delta m_d$, the limit on 
$\Delta m_s$ and sin 2$\beta$ from the measurement of the CP asymmetry in $J/\psi K^0$ decays.
These constraints are shown in Figure \ref{fig:bande} \cite{ref:bello}.
These measurements provide a set of constraints which are obtained by comparing measured 
and expected values of the corresponding quantities,  in the framework of the Standard Model (or 
of any other given model). In practice,  theoretical expressions for these constraints involve several 
additional parameters such as quark masses, decay constants of B mesons and bag-factors. The values of 
these parameters are constrained by other measurements (e.g. the top quark mass) 
or using theoretical expectations.\\
Different statistical methods have been defined to treat experimental and theoretical errors.
The methods essentially differ in the treatment of the latter and can be classified into two main
groups: frequentist and Bayesian. The net result is that, if the same inputs are used, the different
statistical methods select quite similar values for the different CKM parameters \cite{ref:ckmfits}. 
The results in the following are shown using the Bayesian approach.\\ 
Central values and uncertainties taken for the relevant parameters used in 
these analyses are given in Table \ref{tab:inputs} \cite{parodi}.\\
\newpage
The most crucial test is the comparison between the region selected by the measurements
which are sensitive only to the sides of the Unitarity Triangle and the regions selected by 
the direct measurements of the CP violation in the kaon ($\epsilon_K$) or in the B (sin2$\beta$) sector. 
This test is shown in 
Figure \ref{fig:testcp}.

It can be translated quantitatively in the comparison between the value of 
sin2$\beta$ obtained from the measurement of the CP asymmetry in $J/\psi K^0$ decays and the one 
determined from triangle ``sides`` measurements \cite{ref:stocchi},\cite{parodi}:
\begin{eqnarray}
\sin 2 \beta = & 0.725^{+0.055}_{-0.065} & \rm {triangle~sides~ only}     \nonumber \\
\sin 2 \beta = & 0.734 \pm 0.054 & \rm \quad B^0 \rightarrow J/\psi K^0. 
\label{eq:sin2beta}
\end{eqnarray}
The spectacular agreement between these values shows the consistency of the Standard 
Model in describing the CP violation phenomena in terms of one single parameter $\eta$.
It is also an important test of the OPE,HQET and LQCD theories which have been used to extract the
CKM parameters.\\
\newpage
Including all five constraints the results are \cite{ref:stocchi},\cite{parodi}:
\begin{eqnarray}
   \bar {\eta}  =  &  0.357 \pm 0.027              & ~(0.305-0.411)          \nonumber \\ 
   \bar {\rho}  =  &  0.173 \pm 0.046              & ~(0.076-0.260)          \nonumber \\ 
   \sin 2\beta  =  &  0.725 ^{+0.035}_{-0.031}     & ~(0.660-0.789)          \nonumber \\ 
   \sin 2\alpha =  & -0.09 \pm 0.25                & ~(-0.54-0.40)           \nonumber \\ 
   \gamma       =  &  (63.5 \pm 7.0)^{\circ}       & ~(51.0-79.0)^{\circ}    \nonumber \\ 
   \Delta m_s   =  &  (18.0^{+1.7}_{-1.5}) ps^{-1} & ~(15.4-21.7) ps^{-1}.    
\label{eq:allres}
\end{eqnarray}

The ranges within parentheses correspond to 95$\%$ probability.\\
The results on $\Delta m_s$ and $\gamma$ are predictions for those quantities 
which will be measured in  near future.

\section{Conclusions}

\begin{figure}[htb!]
\begin{center}
\epsfig{file=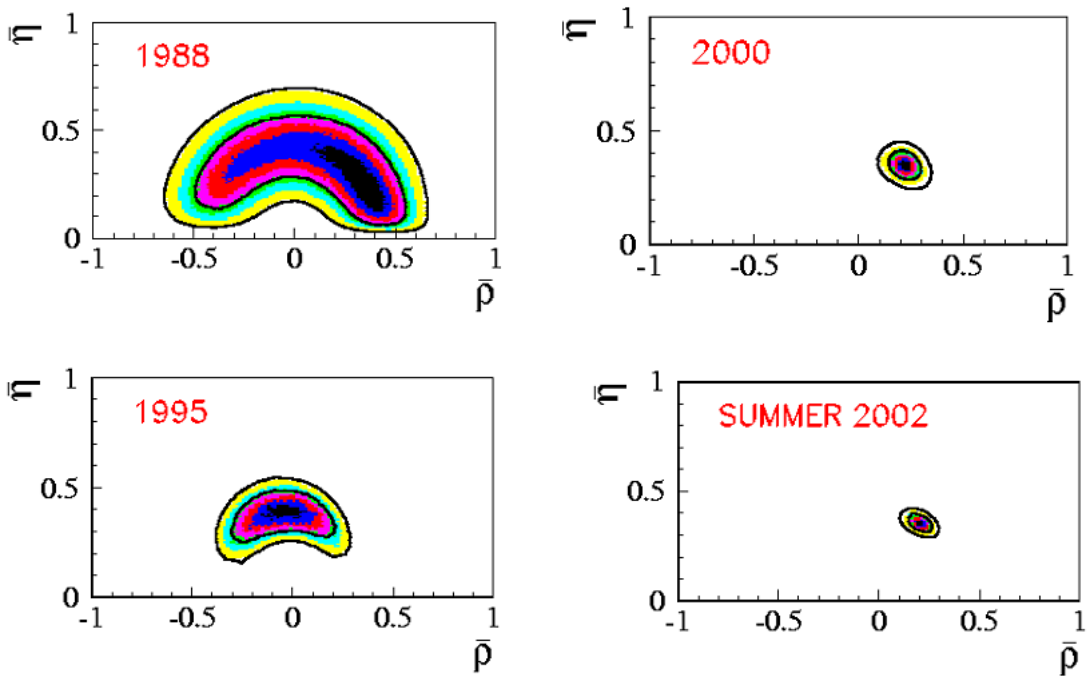,height=9cm}
\caption{ \it {Evolution, over the years, of the allowed regions for $\overline{\rho}$ and 
$\overline{\eta}$ (contours at 68\%, 95\%). }}
\label{fig:storia}
\end{center}
\end{figure}

During the last ten years, our understanding of the flavour sector of the Standard Model
improved. LEP and SLD played a central role.\\
At the start of LEP and SLD, only the $B_d$ and the $B^+$ hadrons were known and 
their properties were under study. Today B hadrons have been carefully
studied and many quantities have already been measured with good precision. 
The hadron lifetimes are now measured at the one/few percent level. 
LEP experiments are the main contributors for the measurement of $|V_{cb}|$, which is known with 
a relative precision better than 2$\%$. In this case, not only, the decay width has 
been measured, but also some of the non-perturbative QCD parameters 
entering in its expression. It is a great experimental
achievement and a success for the theory description of the non-perturbative 
QCD phenomena in the framework of the OPE. \\ 
LEP experiments have been pioneering in determining $|V_{ub}|$ using inclusive methods and reaching
a precision of about 10$\%$, defining a road for future measurements at B-factories.\\
The time behaviour of $B^0-\bar{B^0}$ oscillations has been studied and precisely measured in the 
$B_d^0$ sector. The new and precise measurements performed at the B-Factories confirmed these measurements 
and improved the precision by a factor two. The oscillation frequency $\Delta m_d$ is known with 
a precision of about 1$\%$. $B_s^0-\bar{B_s^0}$ oscillations have not been measured so far,
but this search has pushed the experimental limit on the oscillation frequency $\Delta m_s$ 
well beyond any initial prediction for experimental capabilities. 
SLD experiment has played a central role in this search. 
Today we know that $B_s^0$ oscillates at least 30 times faster than $B_d^0$ mesons. 
The frequency of the $B_s^0-\bar{B_s^0}$ oscillations will be soon measured at the Tevatron. 
Nevertheless the impact of the actual limit on $\Delta m_s$ for the determination of the unitarity 
triangle parameters is crucial.\\
The unitarity triangle parameters are today known within good precision.
The evolution of our knowledge concerning the allowed region in the $\overline{\rho}$-$\overline{\eta}$
plane is shown in Figure \ref{fig:storia}. The reduction in size of the error bands, from the year 1995 
to 2000, is essentially due to the analyses 
described in this paper and to the progress in the OPE, HQET and LQCD theories. 
The reduction between 2000 and 2002 is also driven by the precise measurements of sin 2 $\beta$
at the $b$-factories.\\

A crucial test has been already done: the comparison between the unitarity triangle parameters, as 
determined with quantities sensitive to the 
sides of the unitarity triangle (semileptonic B decays and oscillations), with the measurements of  
CP violation in the kaon ($\epsilon_K$) and in the B (sin2$\beta$) sectors. This agreement tells us that 
the Standard Model is also working in the flavour sector and it is also an important test of the 
OPE,HQET and LQCD theories which have been used to extract the CKM parameters. 
On the other hand, these tests are at about 10$\%$ level accuracy, the current 
and the next facilities can surely push these tests to a 1$\%$ level.

\section{Acknowledgements}
I would like to thank the organisers for the invitation and for having set up a very
interesting topical conference in a stimulating and nice atmosphere during and after the talks.\\
Thanks to all the LEP and SLD members which have made all of it possible ! I would also like to remember the 
important work made from the members of the Heavy Flavour Working Groups who prepared 
a large fraction of the averages quoted in this note. They are all warmly thanked.\\
Thanks to P. Roudeau for the careful reading of the manuscript.

\end{document}